\def\figsize{9.3cm}
\def\figsiz{8.7cm}

\def\rn{}
\def\nn#1 #2{#2. #1}				
\def\nnn#1 #2 #3{#2. #3. #1}			
\def\nnnn#1 #2 #3 #4{#2. #3. #4 #1}		
\def\nnnnn#1 #2 #3 #4 #5{#2. #3. #4 #5. #1}	
\def\dualand{ and\hbox{ }}				
\def\multiand{, and\hbox{ }}				
\def\rf#1;#2;#3;#4;#5 {{\frenchspacing\par\rn#1, #3 {\bf #4}, #5 (#2). \par}}
\def\rg#1;#2;#3;#4;#5;#6 {{\frenchspacing\par\rn#1, #3 {\bf #4}, #5 (#2). \par}}
\def\rfproc#1;#2;#3;#4;#5;#6 {{\frenchspacing\par\rn#1 #2, in {\it #3}, ed. #4 (#5: #6)\par}}
\def\rfprocp#1;#2;#3;#4;#5;#6;#7 {{\frenchspacing\par\rn#1 #2, in {\it #3}, ed. #4 (#5: #6), p#7\par}}

\def\rg#1;#2;#3;#4;#5;#6 {\par\rn#1 #2, {\it #3}, {\bf #4}, #5 (``#6'') \par}
\def\rf#1;#2;#3;#4;#5 {\par\rn#1, {\it #3}, {\bf #4}, #5 (#2)\par}
\def\rfbook#1;#2;#3;#4;#5 {{\frenchspacing\par\rn#1, {\it #3} (#4: #5, #2)\par}}
\def\rfproc#1;#2;#3;#4;#5;#6 {{\frenchspacing\par\rn#1 #2, in {\it #3}, ed. #4 (#5: #6)\par}}
\def\rfprocp#1;#2;#3;#4;#5;#6;#7 {{\frenchspacing\par\rn#1 #2, in {\it #3}, ed. #4 (#5: #6), p#7\par}}
\def\rfprep#1;#2;#3 {{\par\frenchspacing\rn#1, #3 (#2)\par}}
\def\rfprepp#1;#2;#3 {{\par\rn#1 #2, #3\par}}

\def\etal{{\frenchspacing\it et al.}}

\def\eg{{\frenchspacing\it e.g.}}
\def\etc{{\frenchspacing\it etc.}}

\def\beq#1{\begin{equation}\label{#1}}
\def\eeq{\end{equation}}
\def\beqa#1{\begin{eqnarray}\label{#1}}
\def\eeqa{\end{eqnarray}}
\def\eq#1{equation~(\ref{#1})}

\def\fig#1{Figure~\ref{#1}}
\def\Fig#1{Figure~\ref{#1}}

\def\Sec#1{Section~\ref{#1}}
\def\Sec#1{Section~\ref{#1}}

\def\spose#1{\hbox to 0pt{#1\hss}}
\def\simlt{\mathrel{\spose{\lower 3pt\hbox{$\mathchar"218$}}
     \raise 2.0pt\hbox{$\mathchar"13C$}}}
\def\simgt{\mathrel{\spose{\lower 3pt\hbox{$\mathchar"218$}}
     \raise 2.0pt\hbox{$\mathchar"13E$}}}
\def\simpropto{\mathrel{\spose{\lower 3pt\hbox{$\mathchar"218$}}
     \raise 2.0pt\hbox{$\propto$}}}

\def\ed{\end{document}}

\def\beq#1{\begin{equation}\label{#1}}
\def\eeq{\end{equation}}
\def\beqa#1{\begin{eqnarray}\label{#1}}
\def\eeqa{\end{eqnarray}}
\def\eq#1{equation~(\ref{#1})}

\def\a{{\bf a}}

\def\f{{\bf f}}
\def\g{{\bf g}}

\def\j{{\bf j}}

\def\n{{\bf n}}

\def\r{{\bf r}} 
\def\rhat{\widehat{\bf r}}

\def\x{{\bf x}}
\def\y{{\bf y}}

\def\xhat{\widehat{\bf x}}

\def\zhat{\widehat{\bf z}}
\def\A{{\bf A}}

\def\D{{\bf D}}

\def\F{{\bf F}}

\def\M{{\bf M}}
\def\N{{\bf N}}
\def\P{{\bf P}}

\def\vzero{{\bf 0}}

\documentclass[twocolumn,amsmath,nofootinbib]{revtex4} 
\usepackage{amsfonts,amsbsy,epsf} 
\begin{document}






\date{Submitted to {\it Phys.~Rev.~D} August 31 2009, revised June 9 2010, accepted August 7 2010}

\title{Omniscopes: Large Area Telescope Arrays with only N log N Computational Cost}

\author{Max Tegmark}

\address{Dept.~of Physics \& MIT Kavli Institute, Massachusetts Institute of Technology, Cambridge, MA 02139}

\author{Matias Zaldarriaga}

\address{Institute for Advanced Study, Einstein Drive, Princeton, NJ 08540, USA}

\begin{abstract}
We show that the class of antenna layouts for telescope arrays allowing cheap analysis hardware (with 
correlator cost scaling as $N\log N$ rather than $N^2$ with the number of antennas $N$) is encouragingly large,
including not only previously discussed rectangular grids but also arbitrary hierarchies of such grids, 
with arbitrary rotations and shears at each level. 
We show that all correlations for such a 2D array with an $n$-level hierarchy 
can be efficiently computed via a Fast Fourier Transform in not $2$ but $2n$ dimensions.
This can allow major correlator cost reductions for science applications requiring exquisite sensitivity at widely separated angular scales, for example 
21cm tomography (where short baselines are needed to probe 
the cosmological signal and long baselines are needed for point source removal),
helping enable future 21cm experiments with thousands or millions of cheap dipole-like antennas.
Such hierarchical grids combine the angular resolution advantage of traditional array layouts with the cost advantage of
a rectangular Fast Fourier Transform Telescope.
We also describe an algorithm for how a subclass of hierarchical arrays can efficiently use rotation synthesis to 
produce global sky maps with minimal noise and a well-characterized synthesized beam.
\end{abstract}


  
\maketitle



\setcounter{footnote}{0}

\def\thetamin{\theta_{\rm min}}
\def\thetamax{\theta_{\rm max}}

\section{Introduction}
\label{IntroSec}

There is now strong community interest in building more sensitive radio telescopes, stemming from diverse 
science opportunities that range from planets to pulsars, from black holes to cosmology \cite{SKAscienceBook}.
However, greater sensitivity requires greater collecting area, which in turn increases cost.
The cost for a steerable single-dish telescope grows faster than linearly with area, and becomes 
prohibitive beyond a certain point, which has bolstered interest in interferometers.
Interferometer arrays can be made arbitrarily large, but for a generic array layout, the cost unfortunately grows
quadratically with area asymptotically, eventually becoming financially unviable.
The reason for this is that for an array of $N$ antennas, all $N(N-1)/2\simpropto N^2$ pairs of antennas need to be correlated to calculate the so-called visibilities which encode the sky image. 
Thus the cost of the hardware performing these computations scales as $N^2$, dominating 
all other costs of the interferometer (which tend to grow linearly) for sufficiently large $N$.
An array of $N\sim 10^6$ cheap dipole-style antennas has the potential to 
greatly improve constraints on dark matter, dark energy, inflation, reionization, {\etc}
\cite{BarkanaLoeb05,Bowman07,McQuinn06,21cmpars,LoebWyithe08,Barger08}, 
and the $N^2$-component of the cost starts dominating already around $N\sim 10^3$ \cite{MWA}.

To overcome this limitation, two cost-cutting approaches have emerged:
\begin{enumerate}
\item Partitioning the array into ``tiles'' of $M$ antennas apiece, each treated as a single element
as in \cite{LOFAR,MWA,PAPER,21CMA}. This cuts the correlation cost 
from $N^2$ to $(N/M)^2$, at the price of reducing the sky area covered by a factor $M$. 
In other words, the time savings entirely come from throwing away available sky information and thus not needing to compute it.
\item Arranging the antennas into a rectangular grid that can be correlated using Fast Fourier Transforms (FFTs)
\cite{Butler61,May84,Nakajima92,Nakajima93,Peterson06,Chang07,fftt}.
This reduces the computational cost from $N^2$ to $N\log_2 N$, but by lacking long baselines between antennas,
a fully sampled square grid provides much lower resolution than a conventional sparse array.
\end{enumerate}
\begin{figure}[ht]
\centerline{\epsfxsize=\figsize\epsffile{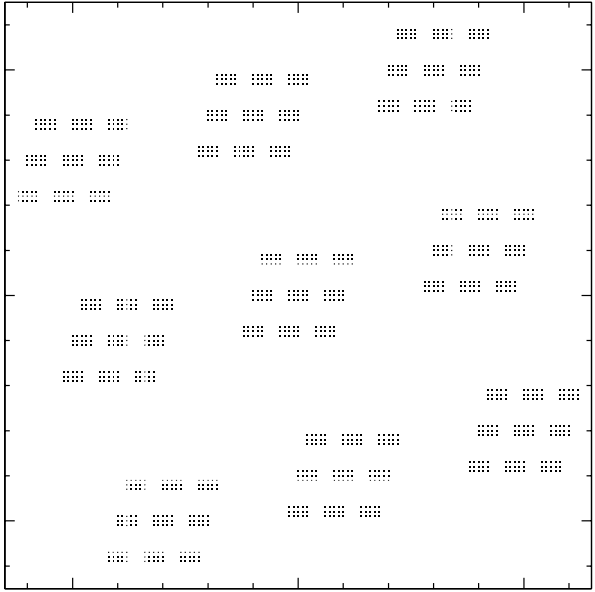}}
\caption{\label{LayoutExampleFig1}
Example of a 3-level hierarchy of antennas, with $5\times 3$ blocks arranged in $3\times 3$ blocks that are in turn 
placed in a $3\times 3$ block. Note that the blocks at each level can be non-square (like at level 1), sheared (like at level 2) 
and rotated (like at level 3), in any combination. We show that this particular array can be efficiently correlated with a 
6-dimensional FFT.
}
\end{figure}

The goal of the present paper is to explore what class of antenna layouts permit cheap ($N\log_2 N$) signal processing
without throwing away any information. 
We will show that this class is encouragingly large,
including not only the above-mentioned rectangular FFT grids, but also arbitrary hierarchies of such grids, 
such as the example shown in \fig{LayoutExampleFig1}. This opens up the possibility of getting the best of both worlds, 
combining affordable signal processing with baseline coverage tailored to specific scientific needs.
We will refer to such an array that is effectively omnidirectional and omnichromatic 
as an {\it omniscope}\footnote{If the electric field is digitized at set of antenna elements with broad primary beams
(like those of the MWA \cite{MWA}, PAPER \cite{PAPER} or 21CMA \cite{21CMA}, say) and these antennas are correlated individually rather 
than in tiles, then the instantaneous field-of-view will be much of the solid angle above the horizon, and the spectral 
coverage in principle covers a large fraction of the range from zero up to the Nyquist frequency,
limited only by antenna and feed response, analog filtering requirements, {\etc}
The designation ``telescope'' feels like a misnomer for such an instrument, since it is not zooming in on a distant object 
subtending a small fraction of the sky. 
}.

The computational savings of this type of arrays comes at a price. By construction, the array instantaneously measures a significantly smaller number
of visibilities (of order $N$ rather than $N^2$). As we will discuss later, this drawback can be compensated at least partially by sky rotation.
Furthermore, this reduced number of instantaneous visibilities implies an enormous redundancy: each visibility is independently measured by order $N$
different antenna pairs, providing an opportunity for vastly improved calibration and systematics control. 

The rest of this paper is organized as follows.
In \Sec{AlgorithmSec}, we describe our hypercube algorithm for analyzing arbitrary multi-level arrays in $N\log_2 N$ time.
In \Sec{ExamplesSec}, we provide examples of how the sky can be probed with arrays in this class.
We summarize our conclusions in \Sec{ConcSec}. In Appendix~\ref{MapmakingSec}, we describe 
how multiple measurements from such arrays can be rapidly merged into sky maps of the full curved sky.

\section{The hypercube algorithm}
\label{AlgorithmSec}

\subsection{Background}

For a modern introduction to radio astronomy, see {\eg} \cite{ThompsonBook}.
A brief and self-contained description of how a rectangular array of $N$ antennas can map essentially the whole sky 
above the horizon at $N\log_2 N$ computational cost can be found in \cite{fftt}, without using any of the approximations that 
are commonly used in radio astronomy. 
One arranges data from each antenna (at a given frequency, say) in a rectangular array, and convolves this array with 
a parity-reversed copy of itself, obtaining an image of what radio astronomers refer to as ``visibilities'' in the ``$uv$ plane''.
Mathematically, this is equivalent to performing an FFT, computing its square modulus, and performing an inverse FFT.
Each visibility is simply the correlation of signals from antennas separated by the corresponding distance vector
(known as a baseline) summed over all pairs with the same separation. 
Such $uv$ plane observations made at different times can then be combined to average down instrumental noise. For any given pointing, only a fraction of pixels in the $uv$ plane will be measured, but Earth's rotation can be exploited to fill in missing 
baselines. This final $uv$ plane image is essentially the Fourier transform of the sky --- \cite{fftt} includes the complications of 
polarization and sky curvature.

\subsection{Hierarchical grids}

A more complicated sparse array layout such as the one in \fig{LayoutExampleFig1} can of course be analyzed with this same 
formalism if the FFTs used for the correlation are performed on a square grid where each antenna falls in a single entry or pixel, but where many of the entries are empty. This would be 
as time consuming as analyzing a non-sparse (fully filled) grid, thus taking no advantage of the sparseness. In the example in \fig{LayoutExampleFig1}, this procedure would increase of the computational cost by a factor of $\sim 8$, and this inefficiency factor would be much larger 
for arrays involving some very long baselines. 

Fortunately, there is a much better way: simply arrange the data from \fig{LayoutExampleFig1} in
a 6-dimensional hypercube of dimensions $5\times 3\times 3\times 3\times 3\times 3$, and perform the correlation in 6 dimensions
using 6-dimensional FFTs. We prove below that this gives the exact same result, and this computation is clearly much faster:
it scales as $N\log_2 N$ where $N$ is not the total number of pixels in \fig{LayoutExampleFig1}, 
but only the number of pixels containing an antenna.

\subsection{A warmup example}
\label{ToyModelSec}

Before getting rigorous, let us consider the following simple toy example to clarify the situation.
Suppose we have a simple one-dimensional array consisting of $N=6$ antennas arranged in 2 groups of 3, located 
along the $x$-axis at positions $x_i=1, 2, 3, 7, 8, 9$ in some units. The corresponding possible separations are all integers  $-8\le\Delta x\le 8$ except $\Delta x=\pm 3$.
Suppose moreover that in some appropriate units, they happen to  measure voltage values $f_i=1, 2, 3, 4, 5, 6$.
Arranging the voltages measured along the x-axis in a vector $\f=(1,2,3,0,0,0,4,5,6)$, we now wish to compute the convolution of $\f$ 
with its parity reversed version $\f^-=(6,5,4,0,0,0,3,2,1)$ to obtain the 
visibility vector $\g\equiv\f\star\f^-$, where $\star$ denotes convolution. This visibilities vector will contain the product of the voltages of antennas separated by a given distance, 
summed over all pairs with the same separation. 
For an $n$-dimensional vector $\f$, this convolution is defined by 
\beq{ConvolutionDefEq}
g_i\equiv (\f\star\f^-)_i \equiv \sum_{j=\max\{1,i+1\}}^{\min\{n,i+n\}} f_j f_{j-i},
\eeq
where $i=-(n-1),...,n-1$ ($i$ corresponds to differences between the integers $j=1,...,n$).

There are now three equivalent ways in which we can compute this. The slowest of all would be to simply compute all the $N^2 /2$ products and sum up the ones corresponding to the same separation. The other two rely on doing FFTs, in either a slower or a faster way. The slower would just perform a zero-padded FFT, squaring the result  and inverse transforming. 
The result is
\beqa{ConvolutionExampleEq1}
\g&=&\left(\hbox{1 2 3 0 0 0 4 5 6}\right)
\star
\left(\hbox{6 5 4 0 0 0 3 2 1}\right)\nonumber\\
&&=\left(\hbox{6 17 32 23 12 0 27 58 91 58 27 0 12 23 32 17 6}\right).\nonumber
\eeqa
However, by instead arranging the measured data in a 2D array, we can obtain the same answer 
by performing a 2D convolution exploiting 2D FFTs:
\beq{ConvolutionExampleEq2}
\left(
\begin{tabular}{rrr}
$1$	&$2$	&$3$\\
$4$	&$5$	&$6$
\end{tabular}
\right)
\star
\left(
\begin{tabular}{rrr}
$6$	&$5$	&$4$\\
$3$	&$2$	&$1$
\end{tabular}
\right)
=
\left(
\begin{tabular}{rrrrr}
$ 6$	&$17$	&$32$	&$23$	&$12$\\
$27$	&$58$	&$91$	&$58$	&$27$\\
$12$	&$23$	&$32$	&$17$	&$ 6$\\
\end{tabular}
\right).
\eeq
The latter is faster, since it eliminates all ``multiply by zero'' steps stemming from the sparseness of the 1D array.
The sparser the array, the greater the speedup. Note that there are two copies of all but
the middle number in the final convolution, which is effectively a palindrome. 
This is of course because the inputs are real-valued; by  performing an real FFT, one avoids computing these numbers twice.

\subsection{The general case}

Let us now investigate why this trick works and generalize it. 
We define a hierarchical grid as a set of points of the form 
\beq{GridEq}
\r = i_1 \a_1 + i_2 \a_2 + ...,
\eeq
where $i_1$, $i_2$, ... are integers spanning some finite ranges and $\a_1$, $\a_2$, ... are vectors. 
We will be mainly interested in the case where these are 2-dimensional vectors, but our algorithm works for any dimensionality.
We will insist that the integer ranges be such that each point $\r$ in the grid corresponds to a unique set of integers.
Our toy example above corresponds to the special case $\a_1=(1,0)$, $\a_2=(6,0)$, $i_1=1,2,3$ and $i_2=0,1$. 
The 3-level hierarchical grid in \fig{LayoutExampleFig1} corresponds to 
$\a_1=(1,0)$, $\a_2=(8,0)$, $\a_3=(40,10)$, $\a_4=(0,1)$, $\a_5=(2,8)$, $\a_6=(-10,40)$,
$n_1=1,...,5$ and $n_2$ through $n_6$ all ranging from $1$ to $3$.

Following the notation from \Sec{ToyModelSec} above, we let $f_\r$ denote the voltage measured by the antenna at position $\r$ (at some time, in some frequency band)
and wish to compute the visibility map $\g\equiv\f\star\f^-$. 
Using \eq{GridEq} and the definition of convolution, we obtain
\beqa{ProofEq1}
g_\r &=& g_{\j + i_1 \a_1 + i_2 \a_2 +...} = \sum_{\r'} f_{\r'}f_{\r'-\r} = \nonumber\\
&=&\sum_{\j' i'_1 i'_2 ...}f_{\j' + i'_1 \a_1 + i'_2 \a_2 +...} f_{(\j' + i'_1 \a_1 + i'_2 \a_2 +...) - (\j + i_1 \a_1 + i_2 \a_2 +...)}\nonumber\\
&=&\sum_{i'_1 i'_2 ...}f_{i'_1 \a_1 + i'_2 \a_2 +...} f_{-\j+(i'_1-i_1)\a_1 + (i'_2-i_2)\a_2 +...}
\eeqa
On the first line, we have included a residual vector $\j\equiv\r-(i_1 \a_1 + i_2 \a_2 +...)$ to allow for the fact that $\r$ may not 
lie on the grid. We have introduced an analogous vector $\j'$ on the second line to ensure that the sum runs over
all $\r'$-values, not merely those in the grid.
Since our measurements $f$ vanish off the grid, we have $f_{\j' + i'_1 \a_1 + i'_2 \a_2 +...}=0$ whenever $\j'\ne\vzero$, 
so that we can set $\j'=\vzero$ and sum only over $i'_1, i'_2,...$.
Rearranging terms on the third line, we see that the second factor now vanishes whenever wherever $\j\ne 0$, which means that 
$g_\r$ equals zero outside the grid and need only be computed on the grid.
Defining 
$F_{i_1 i_2...}\equiv f_{i_1 \a_1 + i_2 \a_2 + ...}$ and 
$G_{i_1 i_2...}\equiv g_{i_1 \a_1 + i_2 \a_2 + ...}$,
we can therefore rewrite \eq{ProofEq1} as 
\beq{ProofEq2}
G_{i_1 i_2...} = \sum_{i'_1 i'_2 ...} F_{i'_1 i'_2} F_{(i'_1-i_1),(i'_2-i_2),...},
\eeq
which we recognize as simply a multidimensional convolution. Note that if an index of $F$ runs from $1$ to $n$, the corresponding index of $G$ runs from $-(n-1)$ to $(n-1)$. 
  
This result is very general: the proof assumes only that the mapping from $\r$ to a set of integers is unique, so grid points 
are even allowed to be interleaved between different levels of the hierarchy as long as they never collide --- 
consider for example $\a_1=(5,0)$, $\a_2=(7,0)$, $i_1=1,..4$ and $i_2=1,...,4$. 
Moreover, the proof is readily generalized to the case where the components of $\r$ are not integers or even rational numbers.

\begin{figure}[ht]
\centerline{\epsfxsize=\figsiz\epsffile{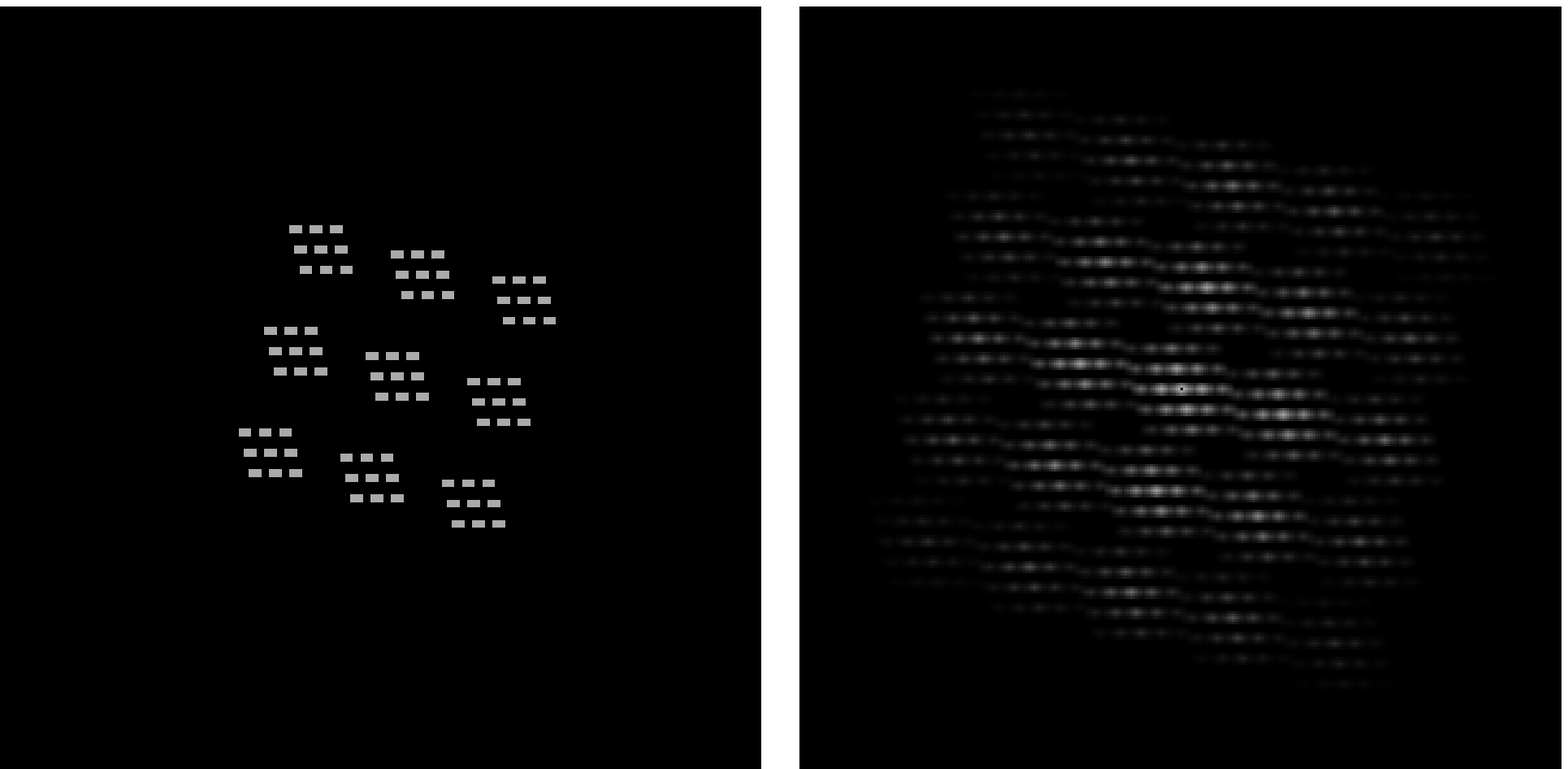}}
\vskip-2.2cm
\centerline{\epsfxsize=\figsiz\epsffile{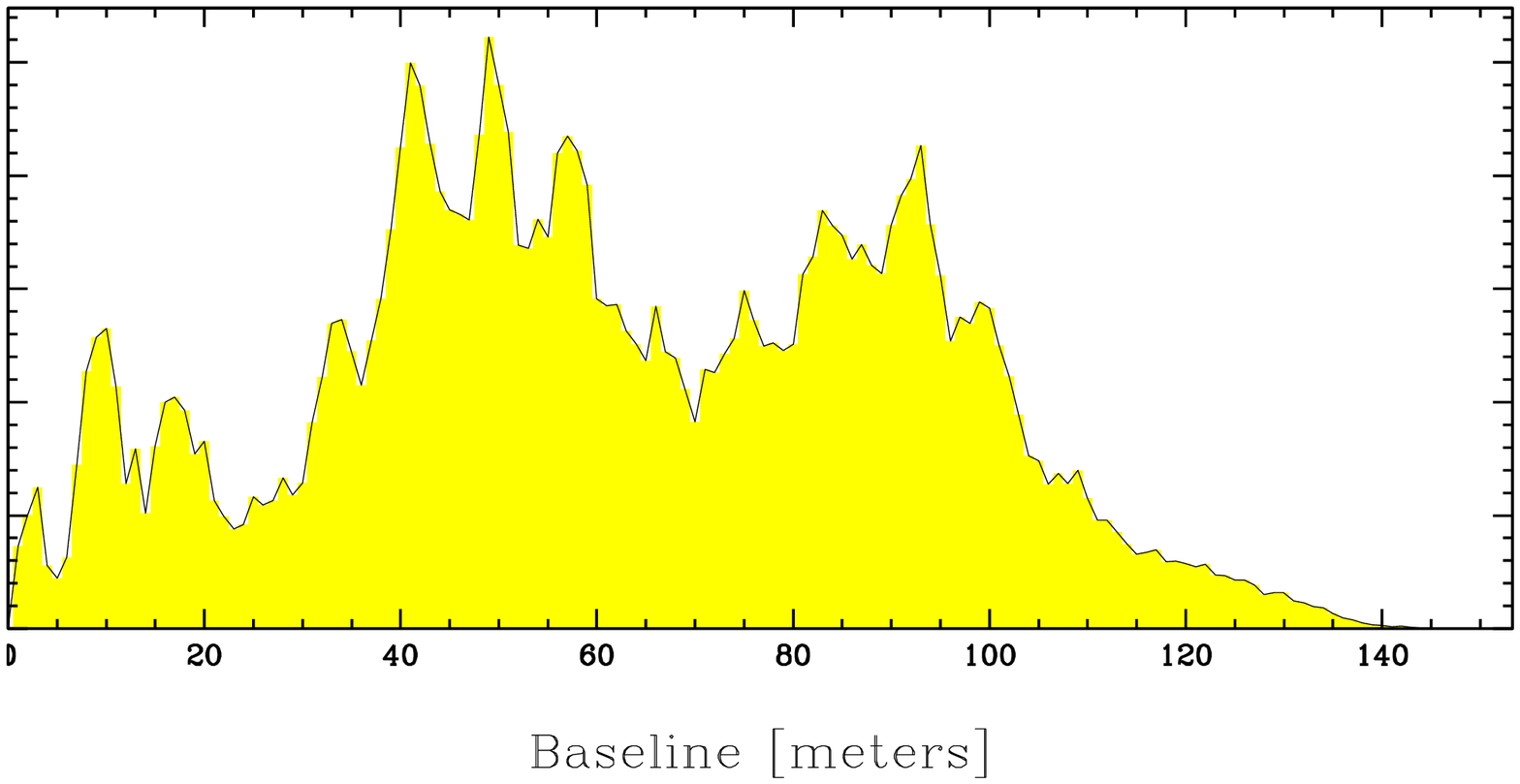}}
\vskip-2.4cm
\caption{\label{ThreeLevelFig}
The antenna layout (top left) from \fig{LayoutExampleFig1} convolved with its parity reversal 
gives the two-dimensional baseline distribution (top right), 
and binning this radially gives the one-dimensional baseline distribution (bottom), which is plotted on a linear
scale. The layout corresponds to
$\a_1=(1,0)$, $\a_2=(8,0)$, $\a_3=(40,10)$, $\a_4=(0,1)$, $\a_5=(2,8)$, $\a_6=(-10,40)$,
$n_1=1,...,5$ and $n_2$ through $n_6$ all ranging from $1$ to $3$. 
}
\end{figure}

\begin{figure}[ht]
\centerline{\epsfxsize=\figsiz\epsffile{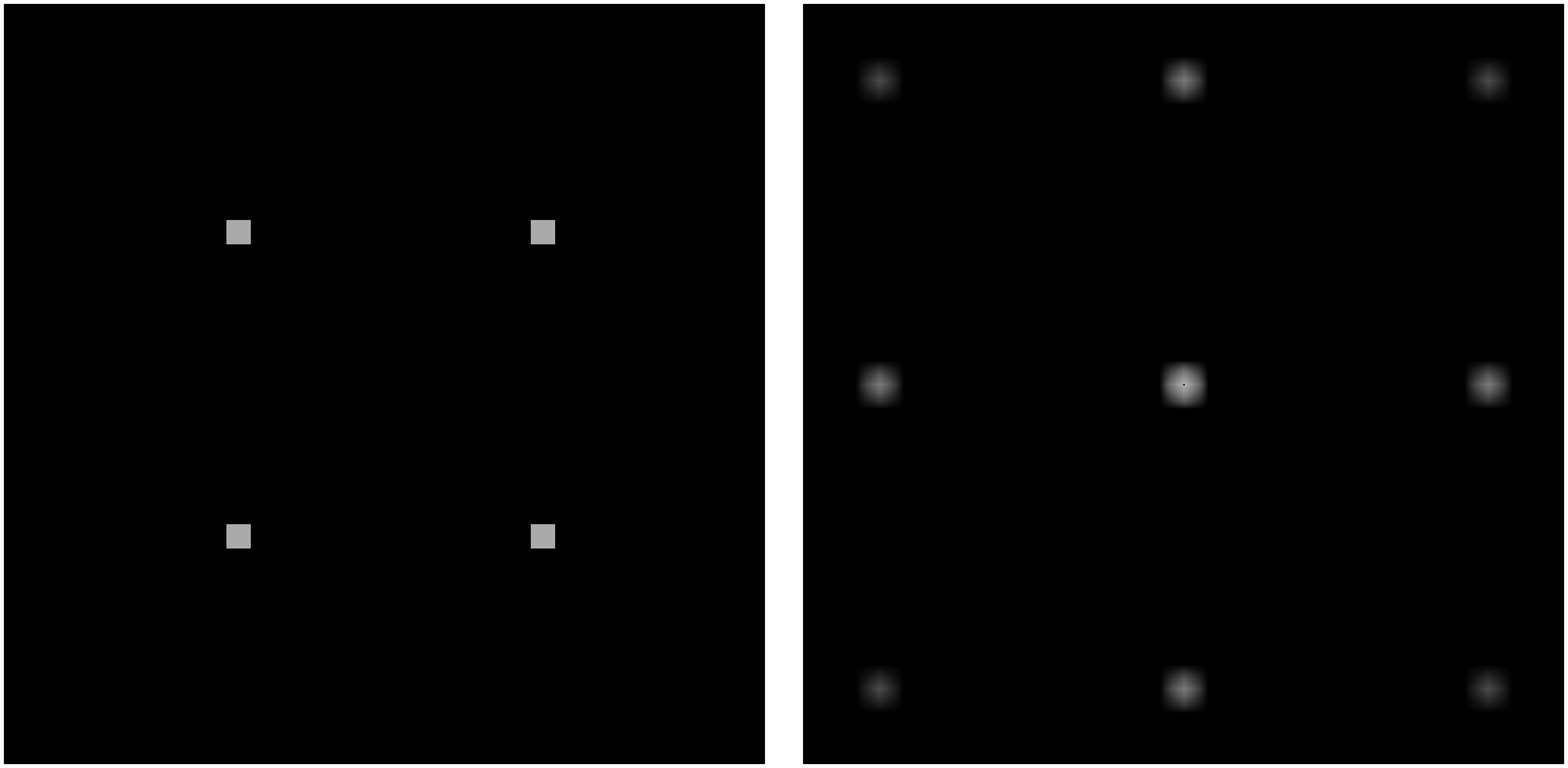}}
\vskip-2.2cm
\centerline{\epsfxsize=\figsiz\epsffile{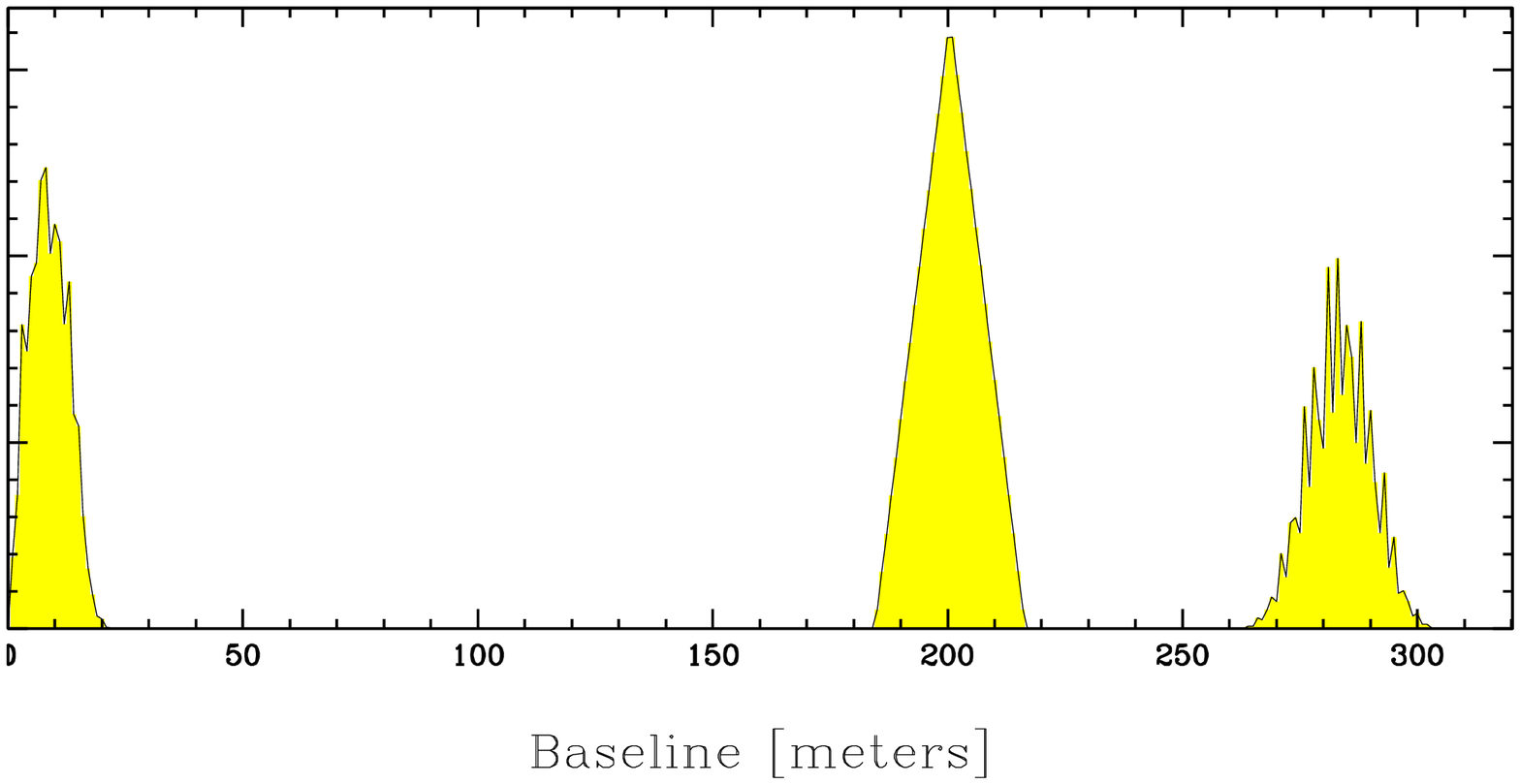}}
\vskip-2.4cm
\caption{\label{BlocksFig}
Antenna layout (top left), 2D baseline distribution (top right) and 1D baseline distribution (bottom)
for the hierarchical ``Blocks'' layout consisting of four widely separated $16\times 16$ antenna blocks
($\a_1=(1,0)$, $\a_2=(200,0)$, $\a_3=(0,1)$, $\a_4=(0,200)$,
$n_1=n_3=16$, $n_2=n_4=2$).
}
\end{figure}

\begin{figure}[ht]
\centerline{\epsfxsize=\figsiz\epsffile{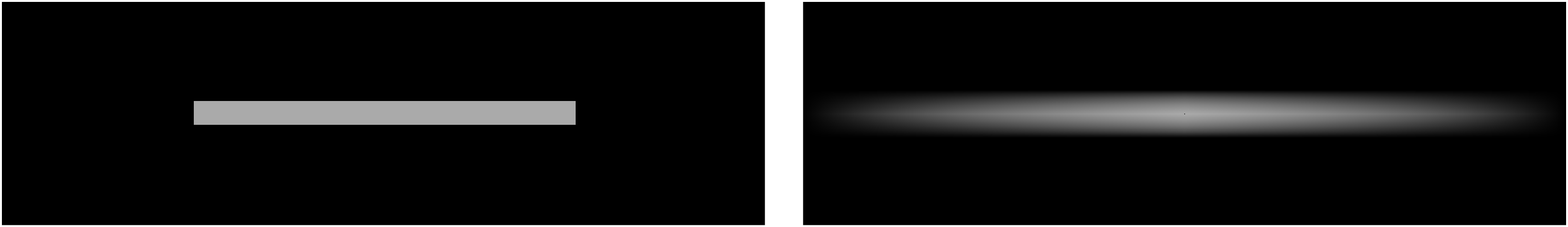}}
\vskip-2.2cm
\centerline{\epsfxsize=\figsiz\epsffile{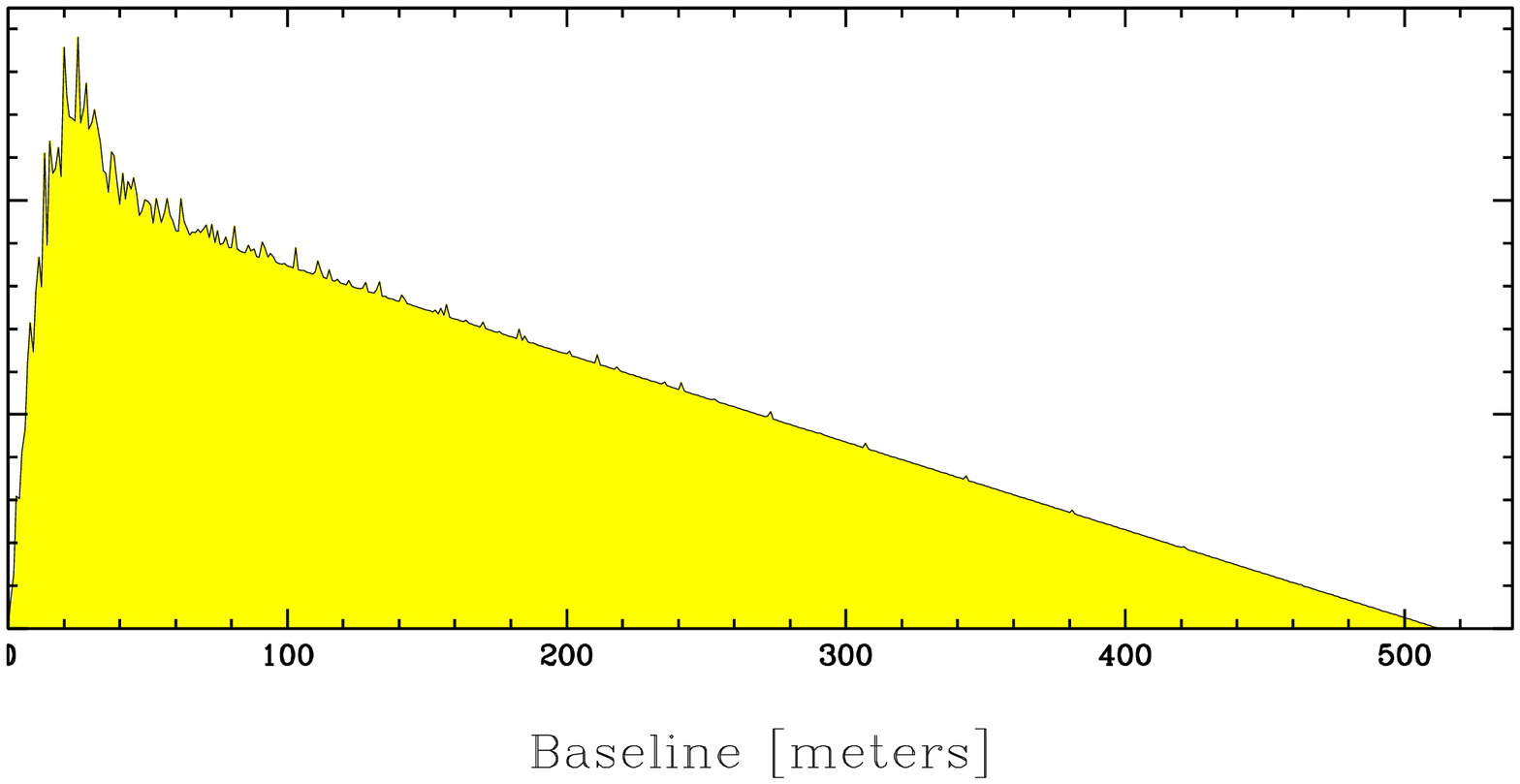}}
\vskip-2.4cm
\caption{\label{PlankFig}
Antenna layout (top left), 2D baseline distribution (top right) and 1D baseline distribution (bottom)
for the ``Plank'' layout consisting of a single $512\times 32$ antenna block.
($\a_1=(1,0)$, $\a_2=(0,1)$, $n_1=512$, $n_2=32$).
}
\end{figure}

\begin{figure}[ht]
\centerline{\epsfxsize=\figsiz\epsffile{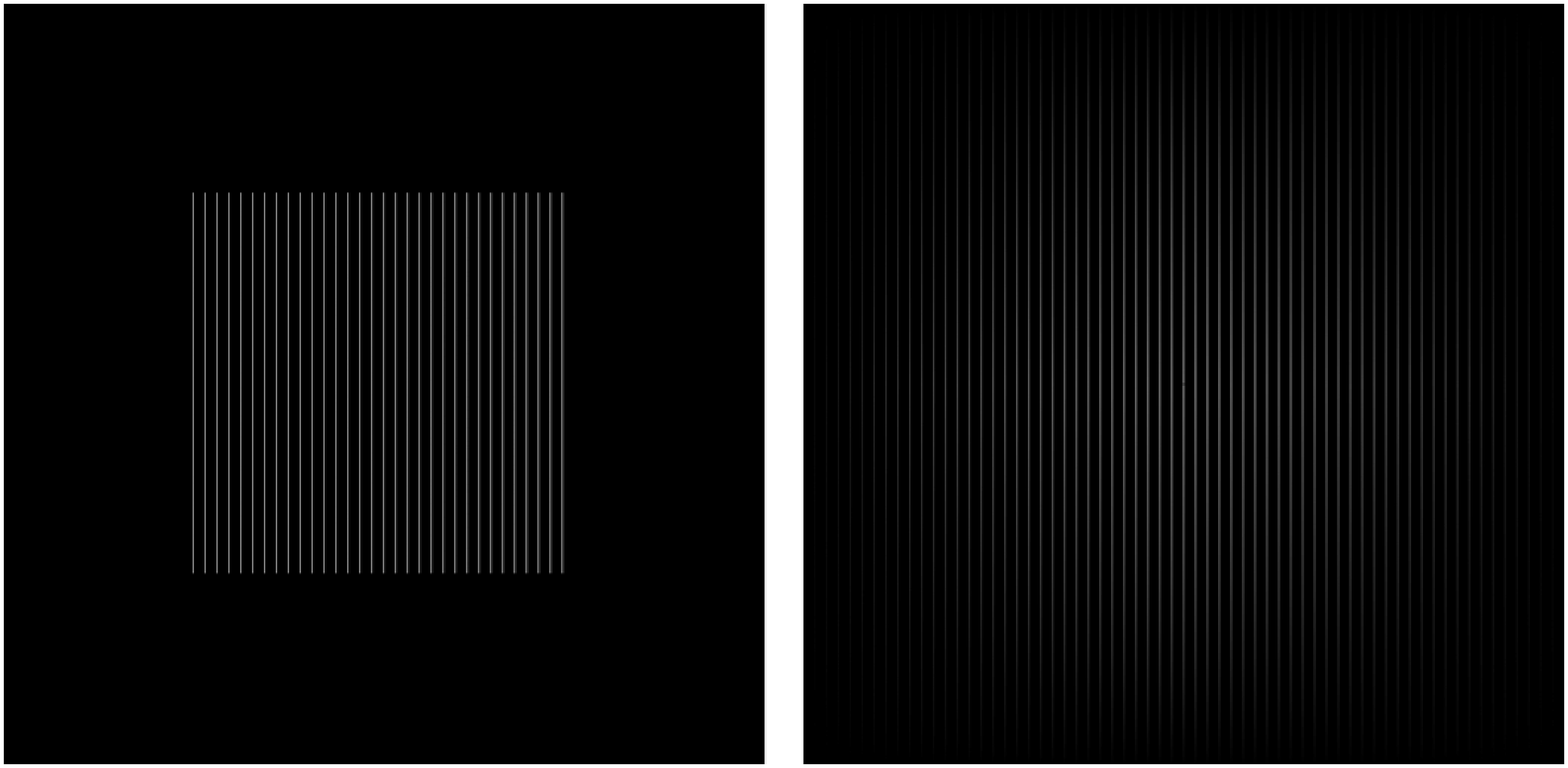}}
\vskip-2.2cm
\centerline{\epsfxsize=\figsiz\epsffile{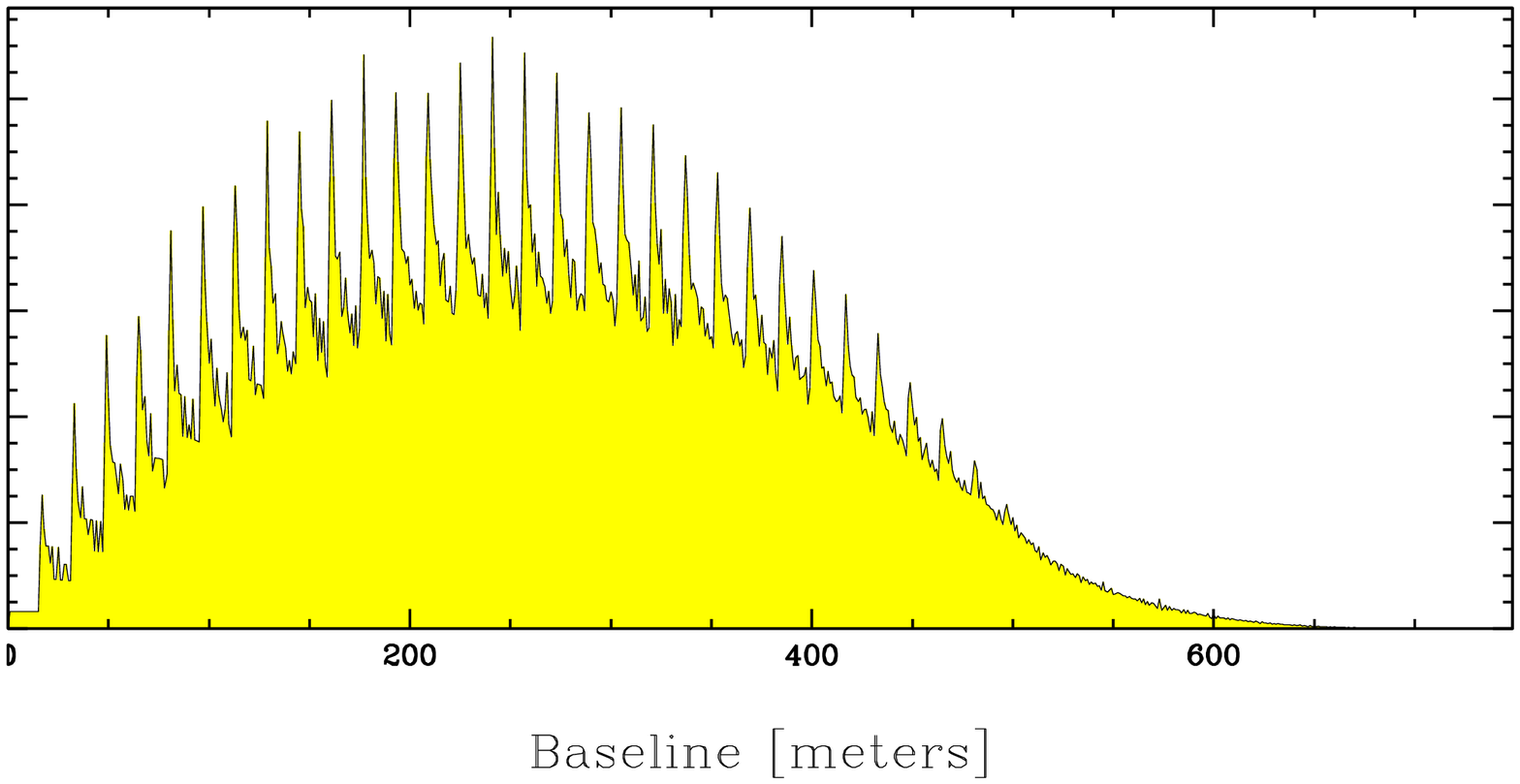}}
\vskip-2.4cm
\caption{\label{StripsFig}
Antenna layout (top left), 2D baseline distribution (top right) and 1D baseline distribution (bottom)
for the ``Strips'' layout consisting of 32 separated rows of antennas
($\a_1=(16,0)$, $\a_2=(0,1)$, $n_1=32$, $n_2=512$).
}
\end{figure}

\begin{figure}[ht]
\centerline{\epsfxsize=\figsiz\epsffile{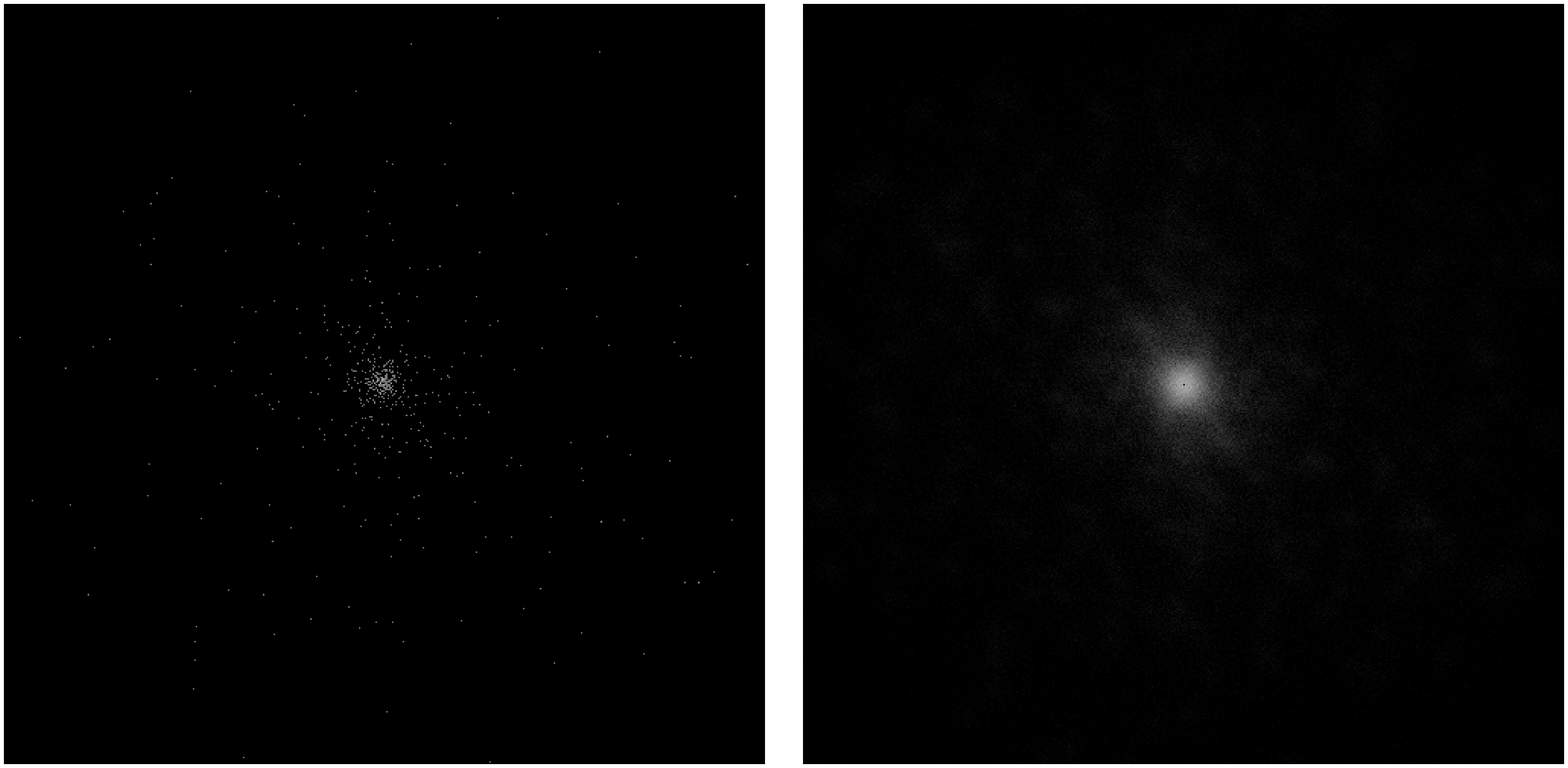}}
\vskip-2.2cm
\centerline{\epsfxsize=\figsiz\epsffile{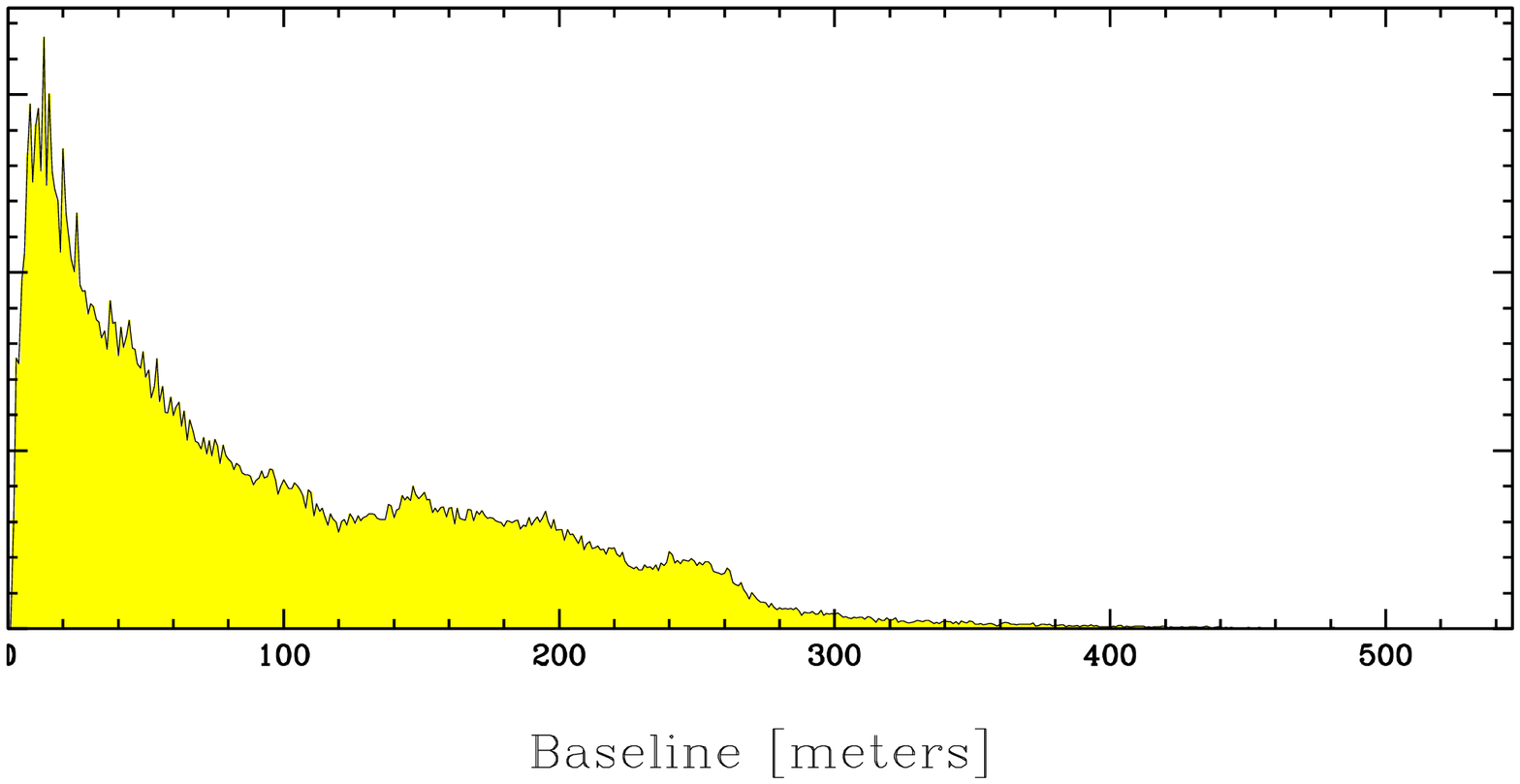}}
\vskip-2.4cm
\caption{\label{rminus2Fig}
Antenna layout (top left), 2D baseline distribution (top right) and 1D baseline distribution (bottom)
for a the simulated MWA design from \cite{Bowman05} with $r^{-2}$ antenna density.
}
\end{figure}

\section{Examples}
\label{ExamplesSec}

Because of the hypercube data analysis algorithm described above, the class of antenna layouts for 
affordable large-area radio telescope arrays is quite large.
In this section, we illustrate the possibilities with some simple examples.

Figures~\ref{ThreeLevelFig}-\ref{StripsFig} show four hierarchical grid omniscope layouts that we nickname
``3-level'' (\fig{ThreeLevelFig}), ``Blocks'' (\fig{BlocksFig}), ``Plank'' (\fig{PlankFig}) and ``Strips'' (\fig{StripsFig}). In all cases, separations are in units of the minimum spacing between antennas, which is our examples is 1 meter.
For comparison, we have also analyzed four non-hierarchical layouts for which the computational cost scales as $N^2$: these are 
500-tile simulations described in \cite{Bowman05} for the MWA-experiment, using random antenna placements with a radial 
density scaling as $r^0$ (uniform), $r^{-1}$, $r^{-2}$ and $r^{-3}$, respectively. The $r^{-2}$ example is illustrated in 2D in \fig{rminus2Fig}.
When plotting the 2D baseline distribution $d$, we convolved the layout map $l$ with its parity reversal, $d=l\star l^{-}$, where the layout map just contains one in every entry. 
We have zeroed the origin and chosen a grey-scale where the intensity is proportional to $\sinh^{-1}(4d/d_{\rm max})$ 
to avoid saturation and allow decent legibility at both small and large values of $d$. Note that the different examples have a different total number of antennas. 

In Appendix~\ref{MapmakingSec}, we will discuss the issue of how these baseline distributions can be filled in by exploiting Earth rotation. 
For our present discussion, we will focus on radially binned versions of these baseline distributions, 
which show how much information is measured on different angular scales.

\begin{figure}[ht]
\centerline{\epsfxsize=\figsiz\epsffile{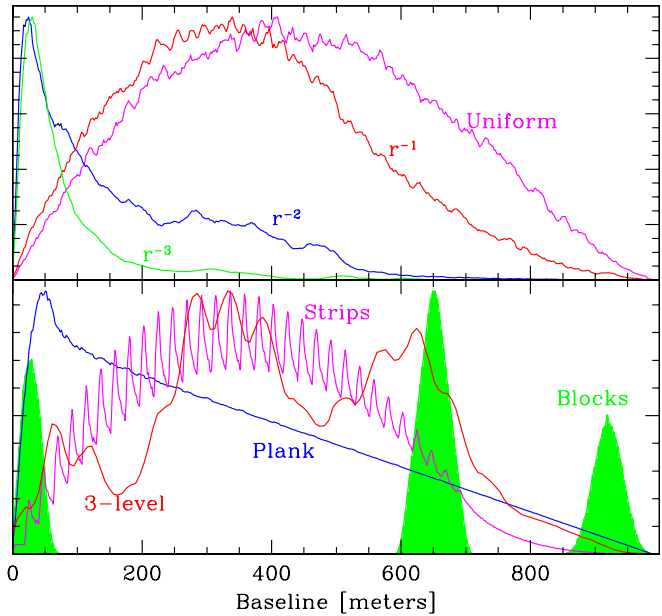}}
\caption{\label{ComparisonFig}
The baseline distributions from four simulated MWA layouts with $N^2$ computational cost  (top) 
are compared with baseline distributions from four hierarchical grids with 
$N\log N$ computational cost (bottom).
For improved legibility, all curves in this figure have been boxchar smoothed with a sliding bin of 
width 5 meters to reduce noise.
}
\end{figure}

The first point illustrated by these examples is that a wide variety baseline distributions can be achieved with a hierarchical
antenna grid --- which is hardly surprising given how large this class of grids is. \Fig{ComparisonFig} compares 
the four hierarchical grid examples (bottom panel) with the four simulated MWA examples (top panel), 
all rescaled to have a 1 km maximum baseline. Because our different examples have different number of antennas, the curves have been arbitrarily rescaled in the vertical direction 
for legibility. Thus only the shapes of the plotted curves are important, not the amplitudes. 
We see that if one has a preference for one of these MWA-distributions, then 
faster-to-analyze hierarchical grids can reproduce at least some of their broad-brush features: note the 
similarities between ``$r^{-2}$ and ``Plank'', and between ``Uniform'' and ``Strips''.

In addition, we see that hierarchical grids can easily offer quite different types of baseline distributions as well. 
For example, ``Blocks'' is seen to provide sensitivity on widely separated angular scales, which may be be desirable for some
science applications (like 21cm tomography, where short baselines are needed to probe 
the cosmological signal and long baselines are needed for point source removal).

 If one drops the requirement of strictly optimal ($N \log N$) computational cost and is willing to stomach an extra factor of two,
 then building a second completely separate omniscope can offer interesting advantages. 
 For example, if two omniscopes with the ``Plank'' layout of \fig{PlankFig} are built rotated by $90^\circ$ relative to one another, 
 then full rotation synthesis can be obtained in merely 6 hours instead of 12, conveniently carried out during a single night. Note that in this way, one computes only 
correlations within each ``Plank" and as a result the one gathers just half the available information.  
Even if daytime observations are acceptable, this greatly improves the $uv$ coverage in generic sky directions from generic latitudes.
In the limit where the width of the Plank shrinks to a single antenna, this is similar in spirit to 
the cylinder telescopes designed by \cite{Peterson06,Chang07}.
(Note that if two cylinders are not parallel (as in the Mills Cross telescope \cite{Burke56}), 
then correlation obviously requires the full $N^2$ cost, since there are $\sim N^2$ separate baselines to compute.)

In the same spirit, another interesting variant \cite{Bunton04,Pen04} is to use 
FFT-correlation separately in a set of $(M/N)$ different rectangular sub-arrays 
(``tiles'') of $M$ antennas each that are translated relative to each other by 
arbitrary amounts, and correlate the corresponding synthesized 
beams between tiles using the brute-force $(N/M)^2$ approach, giving a total computational 
cost of $(M\log M)(N/M)^2\sim (N\log N)(N/M)$.
This is an intermediate case cheaper than the straight $N^2$ and more expensive 
than an $N\log N$ omniscope because it exploits the FFT speedup for two hypercube dimensions, within but not between tiles.

Our examples above assumed no errors in the antenna positions relative to the hierarchical grid.
The question of how antenna position errors affect calibration and image reconstruction 
deserves further work. A first exploration of this problem can be found in \cite{omnical},
showing how calibration errors scale linearly with such position errors, and how this scaling can be improved to quadratic or better with more elaborate software.

\section{Conclusions}
\label{ConcSec}

We have shown that the class of antenna layouts for radio telescope arrays allowing cheap analysis hardware (with correlator
cost scaling as $N\log N$ rather than $N^2$ with the number of antennas $N$) is encouragingly large,
including not only previously discussed rectangular grids but also arbitrary hierarchies of such grids, 
with arbitrary rotations and shears at each level.

This hierarchical grid layout can allow major correlator cost reductions for science 
applications requiring sensitivity at widely separated angular scales, 
for example 21 cm tomography (where short baselines are needed to probe 
the cosmological signal and long baselines are needed for point source removal).
For example, the computers for calculating the $\sim N^2$ correlations constitute (by design) about half the hardware cost 
of the 512-element MWA telescope \cite{MWA}, so for the significantly larger $N$-values that will be needed to perform precision 
21 cm cosmology \cite{21cmpars,LoebWyithe08,Barger08}, computing hardware will completely dominate the budget of any non-hierarchical array.
Looking towards a future with more sensitive instruments, hierarchical grids can therefore be used both to 
cut the costs for a fixed collecting area and to boost the area attainable with a given budget.

If a science goal requires a good approximation to a particular baseline distribution (in 1D or 2D), 
one can use simulated annealing or another suitable tool to solve the nonlinear optimization problem of where to 
place the antennas, either within the hierarchical grid class of layouts or with complete freedom. 
In most cases, general non-hierarchical grids can obviously provide the most accurate approximation, but at a 
much higher price. It is therefore worthwhile to start paying closer attention to what baseline distributions one actually needs
for various science applications, and to quantify whether the case for non-hierarchical layouts is compelling enough to justify the
extra cost. By definition, a hierarchical grid will fill a smaller fraction of the $uv$ plane per snapshot 
(concentrating all its sensitivity measuring or order $N$ rather $N^2$ baselines), but as described in 
Appendix~\ref{MapmakingSec}, this can in many cases be made up for by rotation synthesis. Moreover, the massive $uv$ redundancy, 
whereby most baselines are independently measured by many antenna pairs, offers a powerful 
tool for internal calibration and systematic error control of an omniscope.
In \cite{omnical}, it is shown with detailed simulations that such calibration based on redundant baselines calibration with an omniscope can be made both accurate and computationally feasible.\footnote{Our examples in \Sec{ExamplesSec} assumed no errors in the antenna positions 
relative to the hierarchical grid.
The question of how antenna position errors affect calibration and image reconstruction 
deserves further work. A first exploration of this problem can be found in \cite{omnical},
showing how calibration errors scale linearly with such position errors, and how this 
scaling can be improved to quadratic or better with more elaborate software. 
There are also a number of issues for which the FFT-based correlation does no better and no worse
than traditional $N^2$ correlation; once the baseline distribution is fixed, our hypercube algorithm
is after all just a mathematical trick for computing the exact same numbers as with the $N^2$ method, just faster. Such issues include 
sidelobes due to spatial undersampling, ionospheric modeling 
\cite{Cohen09,Matejek09,Koopmans10}, and direction-dependent gain calibration. The first can be mitigated by planing antennas less than a wavelength apart
at the lowest level of the hierarchical grid. The above-mentioned redundancy can help also with
direction-dependent gain calibration if there are point sources with known positions that completely dominate the signal in their frequency band (like the ORBCOMM satellites), and for the wavelengths most relevant to 21cm tomography, the ionosphere only becomes a problem for baselines above 
the kilometer scale.
}

%


   
  
In summary, there is now strong community interest in building more sensitive radio arrays
consisting of large numbers of cheap dipole-like antennas, and our results indicate a potential for doing so with more bang for the buck.
     
\bigskip
{\bf Acknowledgements:}
The authors wishes to thank Adrian Liu, Andy Lutomirski and anonymous referees for helpful comments, and Adrian Liu and Judd Bowman for providing simulated MWA antenna distributions.
This work was supported by NASA grants NAG5-11099 and NNG 05G40G,
NSF grants AST-0607597, AST-0708534, AST-0908848 and PHY-0855425, 
and fellowships from the David and Lucile
Packard Foundation and the Research Corporation.   

\clearpage

\appendix
\section{Rotation synthesis and mapmaking with hierarchical omniscopes}
\label{MapmakingSec}

Above we described how a hierarchical omniscope could be correlated at an $N\log N$ computational cost per snapshot.
In this Appendix, we will discuss how, in the spirit of the standard radio-astronomy 
technique known as faceting \cite{ThompsonBook}, such snapshots can be combined into a seamless sky map with 
minimal noise and a well-characterized synthesized beam, again at a modest 
computational cost.

\subsection{Optimal mapmaking}

The theory of how to make maps with minimal noise  has been extensively developed and applied in the context of  CMB observations (see for example \cite{mapmaking})\footnote{For a discussion of the related topic of how to go straight from 
snapshots to power spectrum estimates, see \cite{BunnWhite07} 
}. 
Consider a set of $n$ observations  which we arrange in a vector $\y$ (in our case these are the measured correlations, $N(N-1)/2$ visibilities per snapshot in time).  We will assume that $\y$ is of the form
\beq{map}
\y = \A \x+ \n,
\eeq
where $\x$ is an $m$-dimensional vector with the intensity (and polarization) of the sky in different directions,  $\A$ is a known matrix encoding the response of the instrument and $\n$ is the noise. Our goal is to recover $\x$ in a way that does not lose any of the cosmological information. There are many possible linear lossless estimators of $\x$, because if one multiplies any lossless estimator by an invertible matrix, the resulting vector also contains all the available information. As discussed in \cite{mapmaking}, these lossless estimators are of the form
\beq{optimal}
\xhat=\M \A^t \N^{-1} \y,  
\eeq 
where $\M$ is an $m\times m$ matrix. A possible choice of $\M$ is $\M=[\A^t \N^{-1} \A]^{-1}$ which makes the estimator $\x$ unbiased, but there are also other popular choices 
with different attractive features. 
Note that $\xhat$ has size $m$, the size of the pixelized sky map. The  $\y$-vector contains all the observed visibilities and thus in general has a size much greater than $m$. The multiplication by $\A^t$ is the step that accumulates or bins the observations into a single map.   Equation \ref{optimal} constitutes a form of lossless data compression.

\subsection{The key challenge: rapid multiplication by $\A^t$}

For the purpose of our discussion, the crucial point is being able to compute $\xhat$ or a good approximation to it fast, 
in no more than of order $N\log N$ operations which is what is needed to compute all the correlations in $\y$. If we assume that the noise in each visibility is uncorrelated, then
the noise matrix $\N$ is diagonal and multiplying by  $\N^{-1}$ is computationally cheap. 
The key challenge is then being able to apply $\A^t$ to the data in an efficient way. This requires that $\A$ be a sufficiently sparse matrix\footnote{We will assume that one can also multiply by the $\M$-matrix sufficiently fast; this can often be done with an iterative approach once the $\A^t$-multiplication is rapid, as successfully demonstrated by the WMAP team for their microwave background analysis\cite{Hinshaw03}.}. Of course one has the choice of representing the vector $\x$ is any basis, for example one could pixelize the sky in a standard fashion with one pixel per direction on the sky or one could represent the sky by its Fourier modes. The same is true for $\y$.  We will attempt to exploit this freedom to choose a basis in which $\A$ is maximally sparse. 
For simplicity, we will discuss only the unpolarized case below, as polarization does not change the problem in any fundamental way.

\subsubsection{The problem with real space}

The most straightforward choice would be pixelize the sky in angular space, with each entry in $\x$ corresponding to a different direction in the sky.  For a given snapshot, a row in $\A$ then combines the intensity on the sky to give the response of a visibility,  while each row of $\A^t$ combines all the visibilities into an estimate of the sky in a given direction. Defined in this way, $\A^t$ is not sparse or compact because visibilities have response everywhere within the primary beam. 

We can improve the situation by changing basis for $\y$. We could use an FFT to go from the measured visibilities in the $uv$ plane to angle
space (what is usually referred to as the ``dirty map'' in the radio astronomy literature)
in $N\log N$ operations. Now both $\y$ and $\x$ are in angle space. With this choice, the matrix $\A$ is nothing but what is usually
called the instantaneous synthesized beam of the interferometer
(the Fourier transform of the 2D baseline distributions shown in the top right panels of 
Figures~\ref{ThreeLevelFig}-\ref{rminus2Fig}).
Unfortunately, for generic array layouts, 
the instantaneous synthesized
beam tends to be quite complicated, with important side lobe structure, positive and negative regions, {\etc}  
Although better than before, the $\A$
matrix is therefore still not very compact, so multiplying the data by $\A^t$ would dominate the computational costs, spoiling the advantage of
the omniscope design. 

\subsubsection{The problem with Fourier space}

If the omniscope is located at the South Pole, then what to do is rather trivial: one simply aggregates all correlation 
measurements in a single $uv$ plane. That is to say, we use the Fourier  basis to describe $\x$. For each snapshot in time, an observed visibility is the convolution of the Fourier transform of the sky with the Fourier transform of the primary beam. The baselines probed by each antenna pair simply rotates with Earth. In this case, the entries of the $\A$ matrix are given by the Fourier transform of the primary beam, which can be made very compact, so the $\A^t$ multiplication is computationally cheap.  
One accumulates the observations in the $uv$ plane, and after 12 hours, one can then Fourier transform the resulting intensity to recover a map of the Southern half of the 
sky in an easy-to-interpret projection \cite{fftt}. Fourier transforming the corresponding $uv$ weight map recovers the effective synthesized beam.

Unfortunately, for a general  array layout, it is well-known that there are no easy solutions for mapping a 
large solid angle \cite{ThompsonBook}, because of the complication that the plane of the array rotates over time. One cannot simply accumulate observations in the same $uv$ plane.
Although instantaneously the $uv$ plane is clearly the best choice, the $uv$ plane is defined relative to the 
interferometer rather than fixed to the sky, so it is not a good basis to parametrize the sky once long observations are involved. 

\subsubsection{A solution: multiple $uv$ planes}

There is, however, a class of omniscope designs for which the global mapping problem can be easily solved 
for an arbitrary observing location on Earth: the case of a hierarchical grid where the lowest level of the hierarchy
consists of square $n\times n$ antenna blocks that are fully sampled 
(with the individual antennas of order a wavelength apart).
If each of these blocks is operated as a phased array, they can all be digitally pointed to a generic point $\rhat$ in the sky and 
will be sensitive only to signals from within about an angle $\Delta\theta\sim 1/n$ around this point. 
As long as $n\gg 1$, the flat sky approximation can be used for this patch of sky, and one can use standard radio astronomy procedures to
combine all digital pointings towards that 
sky patch in a $uv$ plane defined as the Fourier dual of the tangent plane to $\rhat$ (extended of order $\Delta\theta$ around $\rhat$). 
Note that in this way, the $uv$ plane fixed to the sky rather than to the plane of the instrument, 
and is used to represent a small patch.

With an omniscope consisting of many such $n\times n$ blocks, one can follow exactly the same procedure,  except
that one simultaneously obtains $n^2$ different pointings of the antenna blocks. In practice one would divide the
sky into a number of small patches (``facets'') over which the flat sky approximation is valid and use the Fourier basis to
describe the intensity in each patch. Each of the pointings of the omniscope would then be assigned to the $uv$
plane of one of these patches.

\begin{figure}[ht]
\centerline{\epsfxsize=\figsiz\epsffile{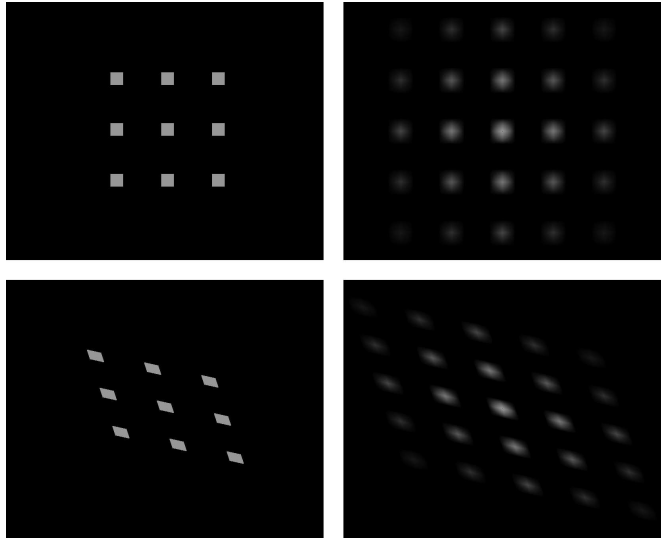}}
\caption{\label{NineBlockFig}
The antenna layout (top left) and 2D baseline distribution (top right) 
of nine $40\times 40$ antenna blocks 
appear deformed (bottom panel) when viewed from a sky direction other than the zenith 
(here the elevation is $30^\circ$ and the azimuth is $63^\circ$).
}
\end{figure}

\begin{figure}[ht]
\vskip-1.2cm
\centerline{\epsfxsize=\figsiz\epsffile{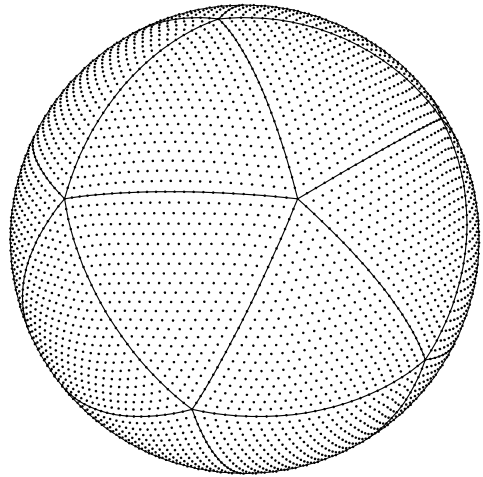}}
\vskip-1.2cm
\caption{\label{IcosahedronFig}
To make the flat sky approximation as accurate as possible within a fixed number of pixels, 
it is helpful to make the pixels hexagonal as in this 
equal-area icosahedron-based scheme \cite{icosahedron} (the pixel centers are shown as dots).
}
\end{figure}

As an illustration, consider the following specific example.
We have nine $n\times n$ antenna blocks with $n=40$, arranged as in \fig{NineBlockFig}.
Since the primary beam size of each such block will be a few degrees, we partition the sky into 6252 hexagonal pixels of roughly this size using the
pixelization method of \cite{icosahedron} as illustrated in 
\fig{IcosahedronFig}: each point in the figure corresponds to a pixel center, and each point (unit vector $\rhat$)
belongs to the pixel whose center it is closest to. This makes generic pixels hexagonal, which has the advantage of minimizing 
sky curvature effects \cite{icosahedron}.
For each pixel center, we keep track of an associated $uv$ plane. Note that the sky pixelization  is fixed in the sense that it does not change as the Earth rotates. 
We now process the omniscope data as follows:
\begin{enumerate}
\item Collect data for say 1 second.
\item FFT the data from each antenna in the time domain and calibrate as necessary. 
\item For a given frequency bin, arrange the data in a hypercube as in \Sec{AlgorithmSec}.
\item Perform a two-dimensional 
FFT for each of the blocks, thereby obtaining digital 
pointings in $32\times 32$ different sky directions $\rhat_i$, $i=1,...,1024$.
\item Perform the correlation in all remaining dimensions using a pair of FFT's as in \Sec{AlgorithmSec}.
\item Merge the data from each such pointing $\rhat_i$ into the $uv$ plane corresponding to the closest pixel center, bearing in mind 
that the baseline distribution must be appropriately rotated and shortened by a factor $\cos\theta\equiv\zhat\cdot\rhat_i$ to reflect the way the 
array looks viewed from the direction $\rhat_i$, and that the polarization vector must be correspondingly rotated. The phases must also be adjusted
to reflect the translation from each pixel center to $\rhat_i$.
For the pointing direction $\rhat_i$ from which the array looks deformed like in \fig{NineBlockFig} (bottom left), the 
corresponding $uv$ weight function thus gets augmented by the corresponding deformed baseline distribution (bottom right).
\item Repeat for all frequency bins of interest.
\item Repeat for as long an integration as desired, exploiting that the pointing directions $\rhat_i$ in celestial coordinates rotate with Earth.
\item FFT the data in each of the 6252 $uv$ planes to obtain maps of the corresponding sky patches, 
and FFT the corresponding $uv$ weight functions to obtain the corresponding synthesized beam functions.
\item Seemlessly merge all the patches into a single map.
\end{enumerate}

It is interesting to compare this approach (for observations in a single $uv$ plane) 
with what \cite{SoftwareHolography} term ``software holography''. 
The similarity is that in both cases, each measured visibility is accumulated in the $uv$-plane not as a delta function 
(by adding the measured complex number to a single $uv$ pixel), but rather as the Fourier transform of the instantaneous 
primary beam (corrected for deformation and translation as in \fig{NineBlockFig}), adding 
in this $uv$ image multiplied by the measured complex number. In both cases, it is important that the $uv$ plane where 
the visibilities and weight functions are accumulated are are well oversampled, to adequately resolve the 
antenna shapes. 
The difference is that software holography does this operation once for every sample (say once every nanosecond), 
whereas with our omniscope approach it is sufficient to do this only once per snapshot (say every second). 
This omniscope approach thus enables major computational savings, 
since all that needs to happen for every sample is the much simpler FFT-processing, where no oversampling is needed and each 
measured number is inserted in only one place (in the hypercube).

Perhaps the only subtlety in the omniscope mapmaking algorithm above relates to the splitting of the sky into domains and the combination of the different domains into a single final map. In particular, a given pointing of the omniscope could land on the boundary between domains. We can treat this by making the domains overlap as follows. First write
\beq{tilesky}
\y= \A \x + \n \equiv \P \F \D \x + \n,
\eeq
where we have decomposed the $\A$ matrix into several pieces. First $\D$ splits the sky into domains. Note that we
can decide to assign each pixel on the sky to more than one domain. For example we can choose to make the domains
overlap in such a way that whenever the center of a pointing of the omniscope falls in one domain, we will assign
the entire primary beam to that domain. As a result, the domains will have to overlap as a pointing whose center is
near the edge of a domain ``spill" into neighboring ones. Rather than assigning measurements of each pointing to
more than one patch, we simply make the patches overlap and assign each pointing to only one patch. As a result the
$\D$-matrix is not a square matrix. If for example we choose to have each pixel on the sky belong to three domains,
$\D$ is a $3m \times m$ matrix. The matrix $\F$ simply changes basis to describe the intensity in each domain by
its Fourier transform, that is to say it goes from angle space to the $uv$ plane.  Finally, each visibility is
given by the convolution of the Fourier modes in a given patch with the primary beam of the tile, encoded in the
matrix $\P$. As described above, the projected separation of each pair of antennas and the location of the center
of the pointing relative to the center of the sky patch are needed to compute $\P$. Note also that the size and
shape of the primary beam also depends on the direction of the pointing. 

Now the estimate of the sky becomes
\beq{optimal2}
\xhat=\M \D^t \F^t  \P^t \N^{-1} \y,  
\eeq 
where $\P^t$ accumulates the visibility measurements in the individual $uv$ planes fixed to the sky. After all the measurements were accumulated in each domain,  $\F^t$ goes back to angle space from Fourier space. Finally, $\D^t$ combines the measurements in all the domains into a single sky map.

\end{document}